\documentclass[useAMS,usenatbib]{mn2e}
\usepackage{graphicx}

\title[Search for pulsating PMS stars in NGC\,6383]{Search for pulsating pre-main sequence stars in NGC\,6383  
\thanks{based on observations with the 0.9m telescope at Cerro Tololo 
Interamerican Observatory (CTIO), La Serena, Chile.}}
\author[K. Zwintz]{K. Zwintz$^{1}$\thanks{E-mail:
zwintz@astro.univie.ac.at}, M. Marconi$^{2}$, P. Reegen$^{1}$, W.W. Weiss$^{1}$\\
$^{1}$Department of Astronomy, University of Vienna, T\"urkenschanzstra\ss e 17, A-1180 Vienna, Austria\\
$^{2}$Osservatorio Astronomico di Capodimonte, Via Moiariello 16, 80131 Napoli, Italy}

\begin{document}

\date{Accepted . Received ; in original form }

\pagerange{\pageref{firstpage}--\pageref{lastpage}} \pubyear{2005}

\maketitle

\label{firstpage}

\begin{abstract}
A search for pulsating pre-main sequence (PMS) stars was performed in the young open cluster NGC 6383 using CCD time series photometry in Johnson {\it B \& V} filters. With an age of only $\sim$1.7 million years all cluster members later than spectral type A0 have not reached the ZAMS yet, hence being ideal candidates for investigating PMS pulsation among A and F type stars. In total 286 stars have been analyzed using classical Fourier techniques. From about a dozen of stars within the boundaries of the classical instability strip, two stars were found to pulsate: NGC 6383 \#170, with five frequencies simultaneously, and NGC 6383 \#198, with a single frequency. In addition, NGC 6383 \#152 is a suspected PMS variable star, but our data remain inconclusive. 
Linear, non-adiabatic models assuming PMS evolutionary phase and purely radial pulsation were calculated for the two new PMS pulsators. NGC 6383 \#170 appears to pulsate radially in third and fifth overtones, while the other three frequencies seem to be of non-radial nature. NGC 6383 \#198 pulsates monoperiodically, most probably in the third radial overtone.\\
Magnitudes and $B-V$ colours were available in the literature for only one third of all stars and we used them for calibrating the remaining.
\end{abstract}

\begin{keywords}
techniques: photometric, stars: pre-main sequence, stars: variables: $\delta$ Scuti, galaxy: open clusters and associations: NGC 6383
\end{keywords}

\section{Introduction}
The study of the first stages in the formation of stars is one of the currently most active research fields in stellar astronomy. Pre-main sequence (PMS) stars lie between the birthline and the zero-age main sequence (ZAMS) in the Hertzsprung-Russell (HR) diagram. They are often characterized by a high degree of activity, strong near- or far-IR excesses and -- in most cases -- emission lines. 
They show photometric and spectroscopic variability on time scales of minutes to years, indicating that stellar activity begins in the earliest phases of stellar evolution. 
During the contraction towards the main sequence intermediate mass PMS stars possess temperatures and luminosities similar to evolved stars in the classical instability strip. This fact suggests that at least part of their activity is due to stellar pulsation. 

The existence of pulsating PMS stars was first suggested by Breger (1972), who discovered in the young open cluster NGC 2264 $\delta$\,Scuti-like pulsation in two PMS stars. Subsequent observations revealed  similar oscillations in several Herbig Ae/Be field stars, e.g. HR 5999 (Kurtz \& Marang 1995), as well as in PMS members of the young open cluster NGC 6823 (Pigulski, Kolackowski \& Kopacki 2000). All known PMS pulsators populate the spectral range between A2 and F5, their periods lie between 18 minutes (Amado et al. 2004) and several hours, and their amplitudes are at millimagnitude level.

The evolutionary tracks for pre- and post main sequence stars intersect such that stars of fundamentally different evolutionary state have the same effective temperature and luminosity (Breger \& Pamyatnykh 1998). Hence, the determination of the evolutionary state of a field star may be ambiguous. Therefore, young open clusters are most suitable to search for pulsating PMS stars. 

\section{The young open cluster NGC 6383}
The open cluster NGC 6383 ($\alpha_{2000} = 17^{\rm h} 34\fm 8, \delta_{2000} = -32\degr 34\arcmin$) is only $\sim$\,1.7Myr old and belongs to the Sgr OB1 association (together with NGC 6611, NGC 6530 and NGC 6531). It has a diameter of $\sim$20 arcminutes and is centered around the bright spectroscopic binary HD 159176.

Eggen (1961) studied the stars in NGC 6383 using photoelectric photometry and found that all stars later than A0 lie above the ZAMS. Th\'e (1965) performed a photographic study in a circular area of $\sim$12 arcminutes radius and could confirm Eggen's results. The core of the cluster was investigated by Fitzgerald et al. (1978) using photoelectric photometry to derive an average colour excess \mbox{$E(B-V) = 0.33 \pm 0.02$}, a cluster distance of \mbox{1.5 $\pm$ 0.2 kpc}, an apparent distance modulus \mbox{$V-M_V$ = 11.90 $\pm$ 0.25} corresponding to \mbox{$(m-M)_0$ = 10.9} (adopted for our analysis) and a cluster age of \mbox{1.7 $\pm$ 0.4 Myr}. The spectral energy distributions of stars in the central part of NGC 6383 were studied by Th\'e et al. (1985) with photoelectric photometry in the Walraven {\it WULBV}, Cousins {\it VRI} and Near-IR {\it JHKLM} photometric systems. Their derived colour excess \mbox{$E(B-V) = 0.3 \pm 0.01$} and distance of \mbox{1.4 $\pm$ 0.15 kpc} agree well with the values by Fitzgerald et al. (1978). Van den Ancker, Th\'e \& de Winter (2000) investigated the central part of this cluster using low-resolution CCD spectroscopy and confirmed that all stars later than A0 lie above the ZAMS. They also report that several cluster members show an infrared excess,  indicative of the presence of circumstellar dust, heated by the central star. They found H$_{\alpha}$ in emission only for their star \#4 and identified it as a new Herbig Ae star. 

All these characteristics make NGC 6383 an ideal target for the search of PMS pulsating stars.

\section{Observations \& data reduction}
CCD photometric time series  in Johnson {\it B \& V} filters were obtained with the 0.9m telescope at the Cerro Tololo Interamerican Observatory (CTIO), Chile, from Aug 11 to Aug 24, 2001, using the 2084 x 2046 SITe CCD chip, which provides a field of view (fov) of $\sim$13 x 13 arcminutes. In total, 53.25 hours photometry could be acquired within 8 clear nights. 

The basic reductions (bias subtraction, flat-fielding) were performed using standard IRAF routines. The Multi Object Multi Frame (MOMF) software developed by Kjeldsen \& Frandsen (1992) was used to extract the photometric signal. MOMF is optimized to analyze photometric time series (i.e. a large number of CCD frames per night) of semi-crowded fields by combining point-spread function fitting and aperture photometry. MOMF determines absolute and relative magnitudes of each star identified on the frames and their corresponding standard deviations. The absolute values are raw, uncorrected, instrumental magnitudes, whereas the relative light curves are determined by subtracting a weighted mean of all stars on the frame. Variable and non-variable, extremely red or blue stars are used to determine the weighted mean, requiring colour-dependent extinction corrections (see Sect.\,\ref{ext}).

\begin{figure}
\centering{\includegraphics[scale=0.4]{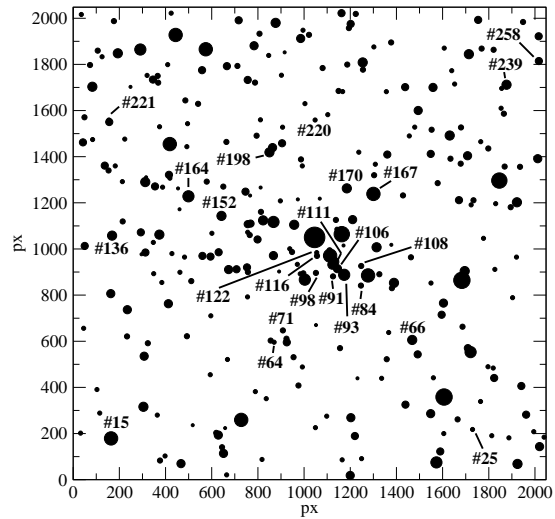}}
\caption{Schematic map of the observed field of NGC 6383 ({\bf fov $\sim$ 13 x 13 arcminutes}, South is at the top and East is to the left) with all stars measured in Johnson {\it B} \& {\it V} in pixels, where 1 px corresponds to 0.33 arcseconds. Identifiers refer to the objects discussed in the text.}
\label{chart}
\end{figure}

286 stars have been identified (see Fig.\,\ref{chart}) for which light curves using the optimum aperture producing minimal point-to-point scatter were generated. Nightly means were subtracted to correct for zero-point changes and  long-term irregular light variations which probably are due to variable extinction by circumstellar dust.

For all 286 stars, a detailed frequency analysis was performed in both filters using the Fourier Analysis program Period98 (Sperl 1998) which is based on the 
Discrete Fourier Transformation (DFT, Deeming 1975) and provides a multi-sine fit option.
A signal was considered to be significant if it exceeds four times the noise level in the amplitude spectrum. (Kuschnig et al. 1997).

Our own star numbers are used, cross references with the literature are given according to the publications by Fitzgerald et al. (1978), e.g. NGC 6383 \#F4, by Th\'e (1965), e.g. NGC 6383 \#T47, and Evans (1978), e.g. NGC 6383 \#EV281. 
%Stars without a number found in the literature are given our own identifiers, e.g. NGC 6383 \#Z25.

\subsection{Colour-dependent extinction}
\label{ext}
A systematic effect was encountered for differential light curves of some stars.
Towards the end of the nights some stars became continuously brighter, but others fainter. The corresponding Bouguer plots (i.e. magnitude vs. airmass) showed that different colours were the explanation. Hence, the extinction correction had to include also the colour-dependent coefficient $k^{\prime}$ 
(Sterken \& Manfroid 1992) in: 
\begin{equation}
m = m_0 - (k^0 + k^{\prime} \cdot CI) \cdot X,
\end{equation}
where $m_0$ is the uncorrected magnitude, $X$ is the airmass and $k^0$ the principal extinction coefficient. In our case, the colour index $CI$ was taken as $(B-V)$. 

\begin{figure}
\centering{\includegraphics[width=0.6\textwidth]{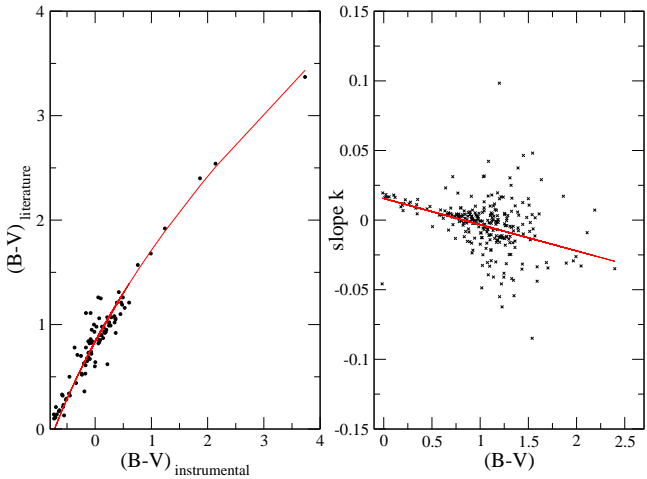}}
\caption{Modelling the colour-dependent extinction effect: {\it left:} Determination of the $(B-V)_{trans}$ of all stars using an inverse second-order polynomial (solid line) which describes the relation between instrumental and literature $(B-V)$ values for 97 of 286 stars. {\it right:} Dependence of the slope of the Bouguer plot, $k$, on $(B-V)_{trans}$ and weighted linear regression. As an example, data from the 8$^{{\rm th}}$ night are shown in this figure, where the slope of the regression is 0.019 $\pm$ 0.010}.
\label{bvk}
\end{figure}

The $(B-V)$ values available in the literature for 97 stars show a clear correlation with the slope, k, of the Bouguer plots (see Fig.\,\ref{bvk}). 
The relation between the 97 stars with $(B-V)$ from the literature and the instrumental $(B-V)$ values from our observations is modelled by an inverse second-order polynomial (solid line in Fig.\,\ref{bvk}). The three polynomial coefficients evaluate to $a_0 = -0.712 \pm 0.052$, $a_1 = +0.692 \pm 0.084$ and $a_2 = +0.177 \pm 0.029$.
$(B-V)_{instr}$ values for all stars could then be transformed to the standard system according to: 
\begin{equation}
(B-V)_{trans} = \sqrt{\frac{a_1^2}{4 \, a_2^2} + \frac{((B-V)_{instr} - a_0)}{a_2}} - \frac{a_1}{2 \, a_2}
\label{poly}
\end{equation}
where $(B-V)_{trans}$ are the transformed indices and \mbox{$(B-V)_{instr}$} are our instrumental values (see Fig.\,\ref{bvk}).

\section{Variable Stars}
About a dozen PMS cluster members have been selected as prime candidates to search for pulsation in NGC 6383 due to their spectral type and/or position in the HR-diagram. The major criterion has been the location of the stars in the region of the classical instability strip.
But only two stars clearly show $\delta$ Scuti-like pulsation within our data, NGC 6383 \#170 ($\alpha_{2000.0} = 17^{\rm h} 34^{\rm m} 37\fs 8$, $\delta_{2000.0} = -32\degr 36\arcmin 19\farcs 5$) and NGC 6383 \#198 ($\alpha_{2000.0} = 17^{\rm h} 34^{\rm m} 48\fs 4$, $\delta_{2000.0} = -32\degr 37\arcmin 21\farcs 6$), and for NGC 6383 \#152 pulsation can only be suspected. 

The other PMS candidates for pulsation remain inconclusive in our data (e.g. they show variability only in one filter or the data quality is not good enough) and therefore need further investigation.

\subsection{Pulsating PMS stars}
\subsubsection{NGC 6383 \#170 (\#F4)}
\begin{figure}
%\vskip 5mm
\centering{\includegraphics[width=0.5\textwidth]{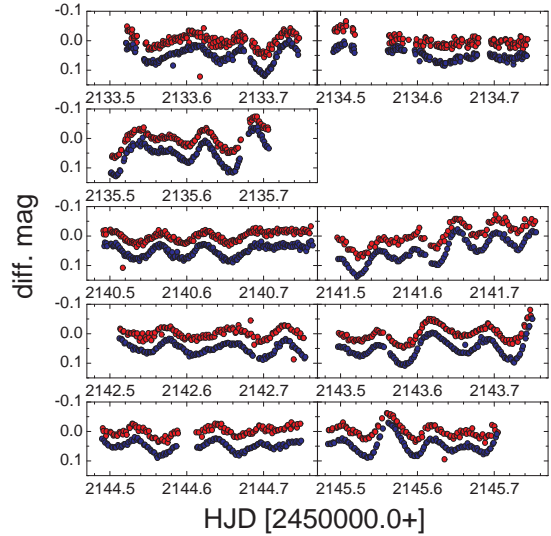}}
\caption{Differential light curves of the pulsating PMS star NGC 6383 \#170: {\it top:} V filter, {\it bottom:} {\it B} filter (shifted for better visibility)}
\label{star4}
\end{figure}

For NGC 6383 \#170 (V\,=\,12.61\,mag), Th\'e, Hageman \& Westerlund (1985) found H$_{\alpha}$ in emission and a large amount of excess radiation in the NIR,  typical for Herbig Ae/Be stars. With a spectral type of A5\,IIIp, a confirmed membership to NGC 6383, and a position above the ZAMS \#170 is a newly discovered PMS pulsator (see Fig.\,\ref{star4}). Five frequencies spanning a period range between 1.24 and 2.89 hours have been detected. These frequencies, Johnson {\it B} \& {\it V} amplitudes and phase shifts are listed in Table\,\ref{fa}. All amplitude and phase errors were computed using the software package {\tt epsim}  (Reegen 2003).

\begin{table}
\begin{center}
\caption{Frequencies, amplitudes and ($\Delta$phase) of phase(V) - phase(B) determined for the two PMS pulsators NGC 6383 \#170 and NGC 6383 \#198, as well as for NGC 6383 \#15, which most probably is a foreground star; the errors in the last digits of the corresponding quantities are given in parentheses.}
\label{fa}
\begin{tabular}{ccrrrr}
\hline
star & no & \multicolumn{1}{c}{frequency} & \multicolumn{1}{c}{{\it V} amp.} & \multicolumn{1}{c}{{\it B} amp.} & \multicolumn{1}{c}{$\Delta$phase} \\
& & \multicolumn{1}{c}{[d$^{-1}$]} & \multicolumn{1}{c}{[mmag]} & \multicolumn{1}{c}{[mmag]} & \\
 \hline
\#170 & f1 & 14.376 & 12.5(8) &	16.0(6)	& -0.243(8) \\
& f2 & 19.436 & 11.3(3) & 14.9(5) 	& -0.092(7) \\
& f3 & 13.766 & 9.8(4) & 12.3(7) 	& 0.474(8) \\
& f4 & 8.295 & 8.6(7) &	11.1(7) 	& 0.035(6) \\
& f5 & 17.653 & 7.6(9) & 9.8(5)		& -0.050(9) \\
\hline
\#198 & f1 & 19.024 & 20.8(9) & 26.4(6) & 0.114(3) \\
\hline
\#15 & f1 & 14.587 & 8.5(3) & 8.4(3) & 0.389(4) \\
& f2 & 16.972 & 4.0(3) & 6.2(2) & -0.605(4) \\
\hline
\end{tabular}
\end{center}
\end{table}

\begin{figure}
\centering{\includegraphics[width=0.35\textwidth]{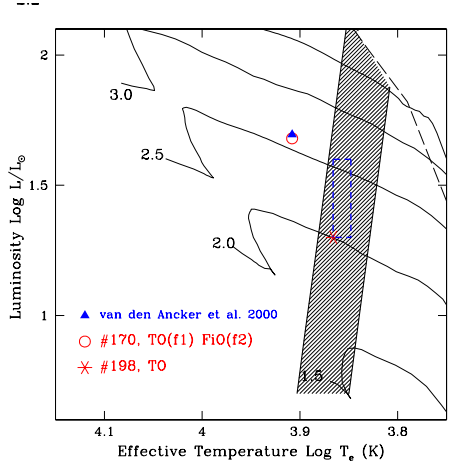}}
\caption{Linear, non-adiabatic radial pulsation models for \mbox{NGC 6383 \#170 and \#198}: solid lines are PMS evolutionary tracks for 1.5, 2.0, 2.5 and 3.0 $M_{\sun}$ (Palla \& Stahler, 1993); the open circle denotes the model for \#170 reproducing the observed frequencies best (see text for detailed explanation), the filled triangle marks the position of the observed values for \#170 derived by van den Ancker, Th\'e and de Winter (2000); the star shows the optimum model for \#198; the dashed box indicates the empirical ranges of log $T_{\rm eff}$ and $\log L/L_{\sun}$ for \#198; the shaded area is the theoretical PMS instability strip for the first three radial modes (Marconi \& Palla, 1998).}
\label{model}
\end{figure}

Linear, non adiabatic pulsation was calculated for radial modes of PMS models  resulting in three possible solutions. No model reproduces all five frequencies simultaneously, but given the probable coexistence of radial and nonradial modes in these stars this could simply mean that not all frequencies correspond to radial pulsation.
The model fitting the observed frequencies best, gives a stellar mass of $2.5 M_{\sun}$, $\log L/L_{\sun} = 1.68$, $T_{\rm eff}=8100 K$, and pulsation in third (f1) and fifth overtones (f2).
The solution seems to be optimal, because it is closest to the parameters  derived spectroscopically by van den Ancker, Th\'e \& de Winter (2000): $\log L/L_{\sun} = 1.69$ and $T_{\rm eff} = 8090 K$ (filled symbol in Fig.\,\ref{model}).

\subsubsection{NGC 6383 \#198 (\#T55)}
\begin{figure}
\centering{\includegraphics[width=0.4\textwidth]{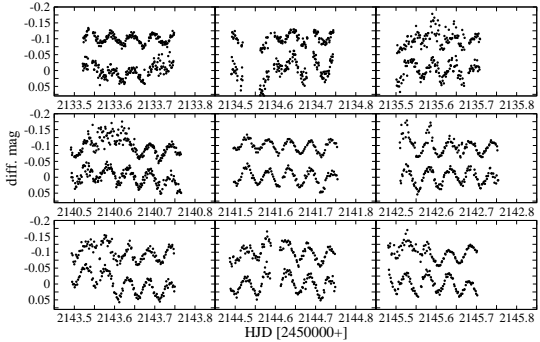}}
\caption{Differential light curves of the PMS pulsator \mbox{NGC 6383 \#198}: {\it top:} {\it V} filter (shifted for better visibility), {\it bottom:} {\it B} filter}
\label{t55}
\end{figure}

Only one frequency of 19.024 d$^{-1}$, corresponding to a period of $\sim$ 1.26 hours, is significant in both filters (see Table\,\ref{fa}), but the light curve shown in Fig.\,\ref{t55} indicates multi-periodicity. 

Calculations of linear, non-adiabatic, radial pulsation models were performed (see Fig.\,\ref{model}). As no spectral classification is available for this star, the $V$ and $(B-V)$ values were used to derive empirical ranges of luminosity and effective temperature based on the transformations given by Kenyon \& Hartmann (1995). These ranges are indicated by the dashed box in Fig.\,\ref{model}. 
Only for pulsation in the third overtone (TO) the theoretical models have temperatures and luminosities close to the observations. Such a high overtone mode is rather difficult to explain in case of monoperiodic pulsation, other modes may be buried in the noise. 

Relying on the cluster membership of NGC 6383 \#198, it seems reasonable that the star pulsates with a single frequency in the third radial overtone having $2.0 M_{\sun}$, $\log L/L_{\sun} = 1.3$ and $T_{\rm eff}=7345 K$. 

\subsubsection{NGC 6383 \#152 (\#T54)}
With $(B-V) = 0.57$ mag \#152 falls within the boundaries of the classical instability strip (Fig.\,\ref{hrd}). But only one significant frequency of 2.55 d$^{-1}$ with a peak-to-peak amplitude of $\sim$30 mmag appears in the {\it B} data. Unfortunately the {\it V} filter data are of worse quality, where the noise is so dominant that no peak in the amplitude spectrum exceeds four times the noise level. 

Although this star has been one of the primary targets in NGC 6383 to search for PMS pulsators, it remains inconclusive in our data. Longer time series of better quality have to be obtained to unambiguously decide about variability.

\subsection{Other Variables}

For several other stars, that probably are not pulsating PMS cluster members, variability could also be detected. A list of suspected variable stars is given in Table\,\ref{susvar}.

\subsubsection{NGC 6383 \#15 (\#T47)}
\begin{figure}
%\vskip 5mm
\centering{\includegraphics[width=0.4\textwidth]{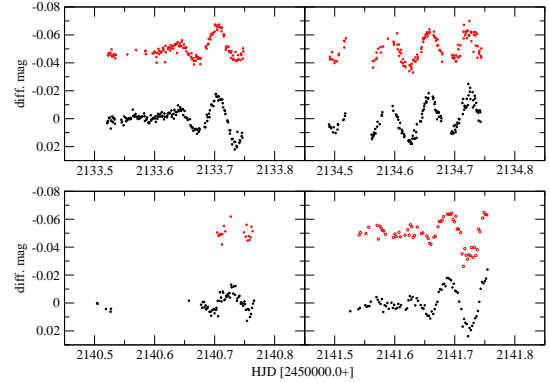}}
\caption{Differential, relative light curves of NGC 6383 \#15: {\it top:} {\it V} filter (shifted for better visibility), {\it bottom:} {\it B} filter}
\label{the47}
\end{figure}

$V = 10.03$ mag and $(B-V) = 0.34$ mag together with its position in the HR-diagram indicate that this star is not a member of the cluster. Two frequencies, corresponding to periods of 1.645 and 1.414 hours, were found to be significant (see Table\,\ref{fa}). 
As it is most likely in the foreground, it seems to be a classical $\delta$ Scuti type star (see Fig.\,\ref{the47}). The amplitude and phase errors were computed using the software package {\tt epsim} (Reegen 2003).

\subsubsection{NGC 6383 \#25}
\begin{figure}
\vskip 10mm
\centering{\includegraphics[scale=0.35]{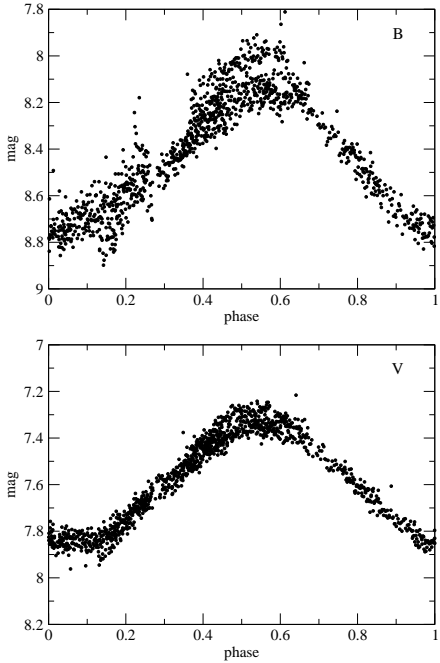}}
\caption{Phase plot of NGC 6383 \#25: {\it top:} {\it B} filter, {\it bottom:} {\it V} filter}
\label{z25ph}
\end{figure}

No astrophysical parameters were available from the literature for NGC 6383 \#25. Our transformation yields $V = 16.77$ mag and $B-V = 1.58$ mag. A frequency of 1.19755 d$^{-1}$, i.e. a period of 20.04 hours, leads to a phase plot shown in Fig.\,\ref{z25ph}. 

$(B-V) = 1.58$ mag corresponds to a spectral type of M5 and is associated to $M/M_{\sun} = 0.21$, and $R/R_{\sun} = 0.27$ according to Schmidt-Kaler (1965). Assuming a rotation period of 20.04 hours, the equatorial rotational velocity is 16.37\,km s$^{\rm -1}$.
This velocity would be in agreement with a rotating, active and weak-lined T Tauri star.

\subsubsection{NGC 6383 \#71}
No astrophysical information was available from the literature for NGC 6383 \#71. Our calculations give $V$ = 15.37 mag and $(B-V)$ = 1.35 mag. If the star belongs to the cluster, it is an early K type star (Schmidt-Kaler 1965). 

\begin{figure}
\vskip 5mm
\centering{\includegraphics[width=0.4\textwidth]{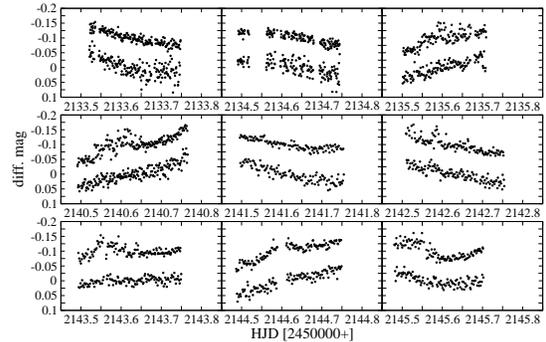}}
\caption{Differential light curves of NGC 6383 \#71: {\it top:} {\it V} filter (shifted for better visibility), {\it bottom:} {\it B} filter}
\label{z71}
\end{figure}

Two frequencies of 2.759 and 2.240 d$^{-1}$, i.e. periods of 8.688 and 10.714 hours, respectively, were detected (Fig. \ref{z71}). A variability on this time scale cannot be explained assuming cluster membership. Permitting NGC 6383 \#71 to be more distant than the cluster itself, interstellar reddening may shift its position in the HR-diagram into the SPB, or $\beta$ Cephei domain. A clear decision can only be drawn from spectroscopy.

\subsubsection{NGC 6383 \#64}
\begin{figure}
\vskip 10mm
\centering{\includegraphics[scale=0.35]{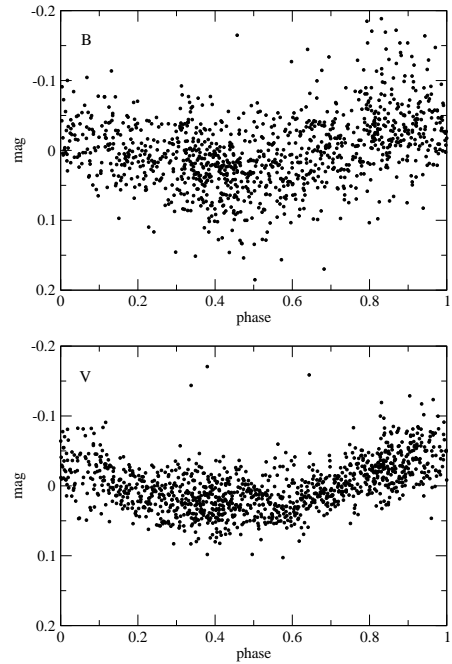}}
\caption{Phase plot of NGC 6383 \#64: {\it top:} {\it B} filter, {\it bottom:} {\it V} filter}
\label{z64ph}
\end{figure}

No information about this star was found in the literature. According to our observations, the star has only 16.72 mag in $V$ and $(B-V) = 1.63$ mag. A single frequency of $\sim$2.499 $d^{-1}$ (corresponding to a period of 9.605 hours) with a peak-to-peak amplitude of approximately 20 mmag is found to be significant in both {\it B} and {\it V} light curves and leads to the phase plot shown in Fig. \ref{z64ph}. 

\begin{table*}
\begin{center}
\caption{Variables and suspected variables in the field of NGC 6383: {\it star} denotes our star number and {\it ref} the cross reference with the literature (according to F ... Fitzgerald et al. (1978), T ... Th\'e (1965)), $M_v$ and $(B-V)_0$ are derived using $E(B-V)=0.33$ mag and $V-M_v = 11.9$ mag, the spectral types ({\it sp}) are taken from the literature}
\label{susvar}
\begin{tabular}{rcccccccl}
\hline
 star & ref & V & B-V & M$_v$ & $(B-V)_0$ & sp & var. in filter & remarks \\ 
 \# & & mag & mag & mag & mag & & B / V / B \& V & \\ 
 \hline
 170 & F 4 & 12.61 & 0.60 & 1.00 & 0.37 & A5 & B \& V & new PMS pulsator\\
 198 & T 55 & 12.90 & 0.36 & 0.93 & 0.3 & - & B \& V & new PMS pulsator\\
 152 & T 54 & 12.45 & 0.70 & 0.44 & 0.24 & - & B \& V & suspected PMS pulsator\\
 15 & T 47 & 10.08 & 0.34 & -1.87 & 0.01 & - & B \& V & foreground $\delta$ Scuti star\\
 25 & - & 16.77 & 1.58 & 4.87 & 1.25 & - & B \& V & T Tauri star\\
 64 & - & 16.72 & 1.63 & 4.82 & 1.30 & - & B \& V & probably not a cluster member\\
 71 & - & 15.37 & 1.35 & 3.47 & 1.02 & - & B \& V & probably not a cluster member\\
\hline
 66 & T 28 & 12.59 & 0.33 & 0.63 & -0.17 & - & B & inconclusive in V\\ 
 84 & - & 15.98 & 1.31 & 4.08 & 0.99 & - & B & inconclusive in V\\ 
 91 & F 10 & 15.3 & 0.9 & 3.41 & 0.62 & - & B & inconclusive in V\\ 
 93 & F 20 & 11.42 & 0.17 & -0.43 & -0.24 & B8 & B \& V  & different periods in B \& V\\ 
 98 & F 11 & 15.10 & 1.06 & 3.26 & 0.81 & - & B \& V & T Tauri candidate\\
 106 & F 6 & 13.77 & 0.52 & 1.93 & 0.27 & A6 & B \& V & known IR excess\\ 
 108 & - & 15.61 & 1.45 & 3.71 & 1.12 & - & B \& V & P$\sim$7.26 hours, inconclusive\\ 
 111 & F 8 & 12.82 & 0.32 & 1.00 & 0.00 & - & B \& V & P$\sim$5.65 hours, inconclusive\\ 
 116 & - & 15.13 & 1.31 & 3.23 & 0.98 & - & B \& V & T Tauri candidate\\
 122 & - & 16.44 & 1.42 & 4.54 & 1.09 & - & B \& V & unresolvable\\ 
 136 & T 52 & 12.33 & 0.72 & 0.43 & 0.39 & - & B & inconclusive in V\\ 
 164 & T 5 & 11.31 & 0.011 & -0.6 & -0.32 & - & B \& V & P$\sim$2 days, amplitude $\sim$ 30 mmag\\
 167 & F 3 & 10.3 & 0.29 & -1.57 & -0.05 & - & B & saturated in V\\ 
 220 & - & 16.67 & 2.03 & 4.77 & 1.70 & - & V & P$\sim$2.87 hours in V\\ 
 221 & - & 16.54 & 1.19 & 4.64 & 0.86 & - & B \& V & irregular variable, T Tauri candidate\\ 
 239 & T 77 & 12.50 & 0.85 & 0.61 & 0.59 & - & B \& V & probably not a cluster member\\
 258 & - & 14.56 & 0.69 & 2.66 & 0.69 & - & B \& V & different periods in B \& V\\ 
\hline
\end{tabular}
\end{center}
\end{table*}

\section{Conclusions}
Photometric time series of 286 stars in the field of the young open cluster NGC 6383 were analyzed using classical Fourier techniques in order to detect pulsation among PMS members of spectral types A to F. A higher number of PMS pulsators would allow to confine the boundaries of the PMS instability strip observationally and especially investigate a possible difference to the classical instability region. 

\begin{figure}
\centering{\includegraphics[scale=0.35]{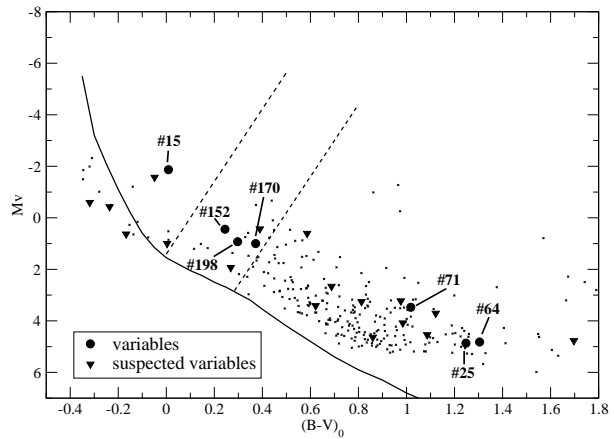}}
\caption{HR-diagram of NGC 6383, the Schmidt-Kaler ZAMS (solid line) and approximate location of the classical instability strip (dashed lines, adopted from Pamyatnykh 2000); small symbols: all analyzed stars, circles: variable stars, triangles: suspected variable stars.}
\label{hrd}
\end{figure}

The computed $V$ and $(B-V)$ values were dereddened using $E(B-V) = 0.33 \pm 0.02$ mag and the apparent distance modulus of $V-M_V = 11.90 \pm 0.25$ corresponding to $(m-M)_0$ = 10.9 and a distance of $1.5 \pm 0.2$ kpc derived by Fitzgerald et al. (1978).
Out of 15 cluster members that fall in the region of the classical instability strip (see Fig.\,\ref{hrd}), for only two, NGC 6383 \#170 and NGC 6383 \#198, pulsation could be clearly detected, whereas for NGC 6383 \#152 variability can only be suspected. This corresponds to $\la$20\% variable stars within the region of the classical instability strip in NGC 6383 down to a noise level of 0.2 mmag in {\it V} and 0.1 mmag in {\it B} in the Fourier domain. 
The percentage of detected pulsating PMS stars in the instability region is somewhat lower than is observed for their post-ZAMS counterparts.

A literature search was performed for additional information about all stars in NGC 6383. For the Herbig Ae star NGC 6383 \#170, one of the two discovered pulsating PMS stars, van den Ancker, Th\'e \& de Winter (2000) found a large infrared excess, H$_{\alpha}$ in emission, and indications for the presence of circumstellar gas in the spectrum. We support the idea of circumstellar material around NGC 6383 \#170, because the raw photometric time series show additional irregular light variations on longer time scales than for pulsation. Only magnitude and colours are available in the literature for NGC 6383 \#198, but its light curve and position in the HR-diagram provide clear evidence for another new PMS pulsator. 

For all stars on the frames, $V$ magnitudes and \makebox{$B-V$} colours are given in Table\,\ref{data} in the Appendix.

\begin{figure}
\centering{\includegraphics[width=0.5\textwidth]{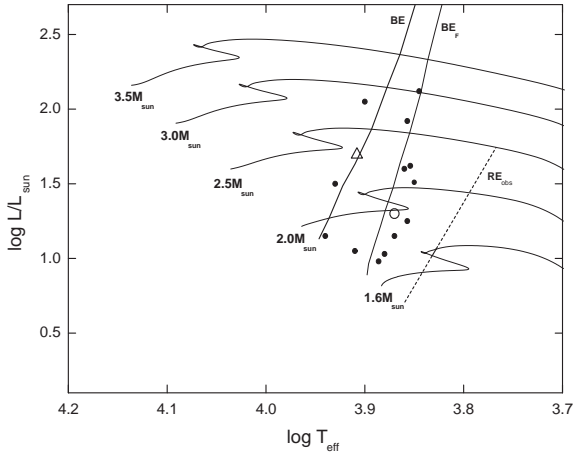}}
\caption{HR-diagram of 15 known PMS pulsators. The open triangle marks the position of NGC 6383 \#170, the open circle the location of NGC 6383 \#198. Also indicated are the PMS evolutionary tracks (D'Antona \& Mazzitelli 1994), the empirical red edge (RE$_{\rm obs}$), the theoretical blue edge for the fundamental mode (BE$_{\rm F}$) and the general theoretical blue edge for the radial overtones (BE) of the classical instability strip. (Breger \& Pamyatnykh 1998)}
\label{pmsstrip}
\end{figure}

The location of 15 known PMS pulsators in the HR-diagram is shown in Fig.\, \ref{pmsstrip}. The values for $T_{\rm eff}$ and $L/L_{\sun}$ for the PMS pulsators are taken from Marconi \& Palla (2003), Marconi (2004, priv. comm.), Amado et al. (2004) and Koen et al. (2003), the PMS evolutionary tracks from D'Antona \& Mazzitelli (1994), the borders of the classical instability strip from Breger \& Pamyatnykh (1998), where RE$_{\rm obs}$ marks the empirical red edge, BE the general blue edge for the radial overtones, and BE$_{\rm F}$ the blue edge for the fundamental mode. However, the number of known PMS pulsators remains insufficient to determine empirically the PMS instability strip. 

\section*{Acknowledgments}

This project was supported by the Austrian {\it Fonds zur F\"orderung der wissenschaftlichen Forschung} (P14984). Fourier Analysis was performed using the program {\sc Period98} written by M. Sperl (1998). Use was made of the WEBDA database, operated at the Institute of Astronomy of the University of Lausanne. Finally, it is a pleasure to acknowledge E. Paunzen for valuable comments on data reduction, A. Pamyatnykh for fruitful discussions, as well as S. Frandsen and T. Arentoft for their patient introduction to the MOMF software.

\appendix

\section[]{Photometric data}

For all stars located on the CCD chip, $V$ and $B-V$ values were computed using our transformation (described in Sect.\,\ref{ext}). Assuming cluster membership they were dereddened using $E(B-V) = 0.33$ mag and $V-M_V = 11.9$ mag (Fitzgerald et al. 1978). The resulting photometric data ($V$, $B-V$, $M_V$ and $(B-V)_0$) are listed in Table\,\ref{data}.

\begin{table*}
 \centering
\caption{Photometric data for all observed stars; cluster membership is assumed; {\it no} lists the star number, {\it ref} denotes the cross reference and star number in the literature, X and Y are coordinates in pixels on the CCD, $M_V$ and $(B-V)_0$ were derived using $E(B-V) = 0.33$ mag and $V-M_V = 11.9$ mag (Fitzgerald et al. 1978).}
\begin{tabular}{@{}rcrrcccc@{}}
  \hline
no & ref & \multicolumn{1}{c}{X} & \multicolumn{1}{c}{Y} & V & $(B-V)$ & $M_V$ & $(B-V)_0$\\
\multicolumn{1}{c}{\#} &   & \multicolumn{1}{c}{[px]} & \multicolumn{1}{c}{[px]} & [mag] & [mag] & [mag] & [mag] \\
\hline
1	&	T 32	&	1199.72	&	18.23	&	13.431	&	0.713	&	1.531	&	0.383	\\
2	&	-	&	664.68	&	20.66	&	16.263	&	1.290	&	4.363	&	0.960	\\
3	&	T 13	&	1924.37	&	68.24	&	12.056	&	0.210	&	0.156	&	-0.120	\\
4	&	T 38	&	466.83	&	69.62	&	13.627	&	0.758	&	1.727	&	0.428	\\
5	&	T 19	&	1573.26	&	74.82	&	11.645	&	1.303	&	-0.255	&	0.973	\\
6	&	-	&	375.99	&	82.95	&	15.563	&	0.878	&	3.663	&	0.548	\\
7	&	-	&	1164.79	&	86.15	&	16.818	&	1.327	&	4.918	&	0.997	\\
8	&	-	&	817.81	&	87.60	&	16.852	&	1.209	&	4.952	&	0.879	\\
9	&	-	&	1248.96	&	91.15	&	16.866	&	1.144	&	4.966	&	0.814	\\
10	&	-	&	397.82	&	102.68	&	16.438	&	1.060	&	4.538	&	0.730	\\
11	&	T 39	&	651.10	&	114.00	&	13.374	&	0.672	&	1.474	&	0.342	\\
12	&	EV 281	&	1589.78	&	122.32	&	14.596	&	1.706	&	2.696	&	1.376	\\
13	&	-	&	644.77	&	140.70	&	15.357	&	1.046	&	3.457	&	0.716	\\
14	&	T 12	&	2019.93	&	143.29	&	13.038	&	0.792	&	1.138	&	0.462	\\
15	&	T 47	&	164.02	&	178.78	&	10.030	&	0.339	&	-1.870	&	0.009	\\
16	&	-	&	1887.08	&	181.27	&	17.003	&	1.250	&	5.103	&	0.920	\\
17	&	-	&	2039.42	&	184.29	&	16.689	&	1.225	&	4.789	&	0.895	\\
18	&	T 31	&	1220.05	&	189.32	&	14.006	&	0.736	&	2.106	&	0.406	\\
19	&	-	&	1812.81	&	190.70	&	16.166	&	1.133	&	4.266	&	0.803	\\
20	&	T 40	&	628.59	&	193.29	&	13.264	&	0.605	&	1.364	&	0.275	\\
21	&	-	&	1605.13	&	199.85	&	16.077	&	1.518	&	4.177	&	1.188	\\
22	&	-	&	31.87	&	201.90	&	16.383	&	0.993	&	4.483	&	0.663	\\
23	&	-	&	619.57	&	202.99	&	15.158	&	1.178	&	3.258	&	0.848	\\
24	&	-	&	1994.76	&	208.01	&	16.794	&	1.219	&	4.894	&	0.889	\\
25	&	-	&	1729.50	&	217.61	&	16.767	&	1.578	&	4.867	&	1.248	\\
26	&	-	&	658.03	&	223.70	&	17.094	&	1.273	&	5.194	&	0.943	\\
27	&	-	&	1049.01	&	225.60	&	16.861	&	1.186	&	4.961	&	0.856	\\
28	&	-	&	517.64	&	236.29	&	17.093	&	2.194	&	5.193	&	1.864	\\
29	&	T 41	&	727.62	&	259.08	&	10.913	&	1.192	&	-0.987	&	0.862	\\
30	&	-	&	1664.79	&	261.29	&	15.050	&	0.843	&	3.150	&	0.513	\\
31	&	T 30	&	1202.61	&	268.73	&	13.379	&	0.914	&	1.479	&	0.584	\\
32	&	-	&	1097.97	&	275.27	&	16.731	&	1.346	&	4.831	&	1.016	\\
33	&	-	&	364.16	&	280.00	&	16.741	&	1.164	&	4.841	&	0.834	\\
34	&	-	&	1961.79	&	281.93	&	14.550	&	1.162	&	2.650	&	0.832	\\
35	&	T 29	&	1548.39	&	286.17	&	13.655	&	0.818	&	1.755	&	0.488	\\
36	&	-	&	115.06	&	288.75	&	16.779	&	1.049	&	4.879	&	0.719	\\
37	&	T 46	&	304.32	&	316.07	&	12.675	&	0.759	&	0.775	&	0.429	\\
38	&	-	&	1439.01	&	325.23	&	14.780	&	0.778	&	2.880	&	0.448	\\
39	&	-	&	1764.32	&	338.96	&	16.081	&	1.105	&	4.181	&	0.775	\\
40	&	-	&	836.25	&	351.06	&	16.542	&	1.428	&	4.642	&	1.098	\\
42	&	-	&	788.92	&	381.86	&	16.293	&	1.240	&	4.393	&	0.910	\\
43	&	-	&	102.83	&	390.66	&	16.382	&	1.196	&	4.482	&	0.866	\\
44	&	-	&	1941.02	&	405.92	&	14.340	&	0.825	&	2.440	&	0.495	\\
45	&	-	&	975.65	&	408.16	&	15.494	&	1.073	&	3.594	&	0.743	\\
46	&	-	&	1231.15	&	439.34	&	17.018	&	1.304	&	5.118	&	0.974	\\
47	&	EV 109	&	1335.71	&	439.39	&	16.672	&	1.114	&	4.772	&	0.784	\\
48	&	-	&	1823.24	&	440.78	&	14.399	&	0.857	&	2.499	&	0.527	\\
49	&	-	&	1559.90	&	441.50	&	16.519	&	1.588	&	4.619	&	1.258	\\
50	&	-	&	593.61	&	454.80	&	16.441	&	1.152	&	4.541	&	0.822	\\
51	&	-	&	1819.27	&	483.77	&	16.148	&	1.062	&	4.248	&	0.732	\\
52	&	-	&	992.62	&	488.38	&	16.745	&	1.425	&	4.845	&	1.095	\\
53	&	-	&	1796.40	&	489.68	&	16.211	&	1.112	&	4.311	&	0.782	\\
54	&	-	&	668.14	&	520.89	&	16.908	&	1.558	&	5.008	&	1.228	\\
55	&	EV 108	&	1357.83	&	522.08	&	15.690	&	1.204	&	3.790	&	0.874	\\
56	&	-	&	954.27	&	530.73	&	15.653	&	1.136	&	3.753	&	0.806	\\
57	&	T 45	&	307.36	&	535.32	&	13.276	&	2.410	&	1.376	&	2.080	\\
\hline
\multicolumn{8}{l}{cross references according to: T ... Th\'e (1965), EV ... Evans (1978),} \\
\multicolumn{8}{l}{F ... Fitzgerald et al. (1978)} \\
\label{data}
\end{tabular}
\end{table*}

\begin{table*}
 \centering
\contcaption{Photometric data}
\begin{tabular}{@{}rcrrcccc@{}}
  \hline
no & ref & \multicolumn{1}{c}{X} & \multicolumn{1}{c}{Y} & V & $(B-V)$ & $M_V$ & $(B-V)_0$\\
\multicolumn{1}{c}{\#} &   & \multicolumn{1}{c}{[px]} & \multicolumn{1}{c}{[px]} & [mag] & [mag] & [mag] & [mag] \\
\hline
58	&	EV 107	&	1492.03	&	543.14	&	14.841	&	1.002	&	2.941	&	0.672	\\
59	&	T 96	&	1721.10	&	553.00	&	11.403	&	0.704	&	-0.497	&	0.374	\\
60	&	EV 111	&	1155.64	&	570.06	&	15.681	&	0.915	&	3.781	&	0.585	\\
61	&	-	&	1709.27	&	570.10	&	14.552	&	0.920	&	2.652	&	0.590	\\
62	&	EV 113	&	323.40	&	591.39	&	15.427	&	1.248	&	3.527	&	0.918	\\
63	&	EV 380	&	925.06	&	595.27	&	14.190	&	0.667	&	2.290	&	0.337	\\
64	&	-	&	870.31	&	595.94	&	16.721	&	1.634	&	4.821	&	1.304	\\
65	&	-	&	855.11	&	602.99	&	15.481	&	1.335	&	3.581	&	1.005	\\
66	&	T 28	&	1468.18	&	605.70	&	12.532	&	0.163	&	0.632	&	-0.167	\\
67	&	-	&	923.89	&	611.24	&	15.374	&	1.115	&	3.474	&	0.785	\\
68	&	-	&	233.11	&	621.07	&	15.116	&	0.982	&	3.216	&	0.652	\\
69	&	EV 114	&	492.68	&	621.79	&	15.942	&	1.034	&	4.042	&	0.704	\\
70	&	EV 110	&	1366.40	&	637.78	&	16.758	&	1.186	&	4.858	&	0.856	\\
71	&	-	&	908.28	&	646.88	&	15.373	&	1.348	&	3.473	&	1.018	\\
72	&	-	&	45.21	&	656.10	&	16.907	&	1.204	&	5.007	&	0.874	\\
73	&	-	&	1686.12	&	662.89	&	15.174	&	0.926	&	3.274	&	0.596	\\
74	&	EV 112	&	1051.58	&	670.23	&	17.131	&	1.262	&	5.231	&	0.932	\\
75	&	EV 115	&	595.97	&	710.18	&	16.084	&	1.091	&	4.184	&	0.761	\\
76	&	-	&	1595.67	&	714.97	&	14.665	&	0.891	&	2.765	&	0.561	\\
77	&	T 44	&	234.60	&	736.71	&	13.560	&	0.706	&	1.660	&	0.376	\\
78	&	T 43	&	412.39	&	762.81	&	13.103	&	0.445	&	1.203	&	0.115	\\
79	&	EV 158	&	1603.79	&	765.28	&	13.797	&	0.691	&	1.897	&	0.361	\\
80	&	-	&	1903.06	&	788.90	&	16.290	&	1.571	&	4.390	&	1.241	\\
81	&	-	&	755.16	&	792.04	&	16.050	&	1.335	&	4.150	&	1.005	\\
82	&	T 50	&	162.81	&	806.29	&	13.527	&	0.943	&	1.627	&	0.613	\\
83	&	-	&	1380.35	&	830.01	&	15.077	&	1.358	&	3.177	&	1.028	\\
84	&	-	&	1246.26	&	841.33	&	15.982	&	1.315	&	4.082	&	0.985	\\
85	&	-	&	1549.76	&	852.24	&	16.491	&	1.283	&	4.591	&	0.953	\\
86	&	T 17	&	1389.39	&	853.14	&	12.649	&	0.659	&	0.749	&	0.329	\\
87	&	-	&	383.82	&	854.74	&	16.256	&	1.213	&	4.356	&	0.883	\\
88	&	-	&	511.36	&	860.75	&	15.543	&	1.114	&	3.643	&	0.784	\\
90	&	F 21	&	1003.22	&	867.06	&	12.000	&	0.773	&	0.100	&	0.443	\\
91	&	F 10	&	1125.23	&	881.18	&	15.312	&	0.952	&	3.412	&	0.622	\\
92	&	F 2	&	1277.07	&	885.23	&	10.403	&	-0.017	&	-1.497	&	-0.347	\\
93	&	F 20	&	1173.62	&	888.08	&	11.473	&	0.094	&	-0.427	&	-0.236	\\
94	&	-	&	1325.99	&	890.34	&	15.870	&	1.013	&	3.970	&	0.683	\\
95	&	-	&	983.26	&	892.09	&	16.770	&	1.742	&	4.870	&	1.412	\\
96	&	-	&	348.81	&	893.05	&	16.332	&	1.133	&	4.432	&	0.803	\\
97	&	-	&	996.66	&	894.20	&	15.576	&	0.955	&	3.676	&	0.625	\\
98	&	F 11	&	1051.19	&	896.15	&	15.155	&	1.143	&	3.255	&	0.813	\\
99	&	-	&	462.23	&	898.94	&	16.814	&	1.409	&	4.914	&	1.079	\\
100	&	-	&	756.70	&	899.89	&	15.735	&	1.348	&	3.835	&	1.018	\\
101	&	-	&	891.73	&	902.12	&	17.578	&	1.650	&	5.678	&	1.320	\\
102	&	T 7	&	1695.73	&	904.24	&	12.752	&	0.339	&	0.852	&	0.009	\\
103	&	T 42	&	673.15	&	910.60	&	13.660	&	0.798	&	1.760	&	0.468	\\
104	&	-	&	1826.27	&	911.37	&	16.469	&	1.250	&	4.569	&	0.920	\\
105	&	-	&	707.38	&	911.91	&	14.043	&	1.142	&	2.143	&	0.812	\\
106	&	F 6	&	1144.36	&	914.89	&	13.833	&	0.598	&	1.933	&	0.268	\\
107	&	-	&	752.96	&	920.26	&	14.324	&	0.852	&	2.424	&	0.522	\\
108	&	-	&	1247.24	&	926.33	&	15.608	&	1.452	&	3.708	&	1.122	\\
109	&	F 7	&	1122.63	&	931.87	&	12.662	&	0.255	&	0.762	&	-0.075	\\
110	&	-	&	970.63	&	933.44	&	16.568	&	1.441	&	4.668	&	1.111	\\
111	&	F 8	&	1130.81	&	944.62	&	12.900	&	0.333	&	1.000	&	0.003	\\
112	&	-	&	1462.80	&	963.98	&	15.573	&	1.100	&	3.673	&	0.770	\\
113	&	-	&	482.24	&	964.66	&	16.406	&	1.044	&	4.506	&	0.714	\\
114	&	EV 132	&	1920.00	&	964.68	&	16.575	&	2.313	&	4.675	&	1.983	\\
115	&	-	&	594.66	&	967.04	&	14.272	&	0.987	&	2.372	&	0.657	\\
116	&	-	&	1055.88	&	969.96	&	15.128	&	1.306	&	3.228	&	0.976	\\
\hline
\multicolumn{8}{l}{cross references according to: T ... Th\'e (1965), EV ... Evans (1978),} \\
\multicolumn{8}{l}{F ... Fitzgerald et al. (1978)} \\
\end{tabular}
\end{table*}

\begin{table*}
 \centering
 \contcaption{Photometric data}
\begin{tabular}{@{}rcrrcccc@{}}
  \hline
no & ref & \multicolumn{1}{c}{X} & \multicolumn{1}{c}{Y} & V & $(B-V)$ & $M_V$ & $(B-V)_0$\\
\multicolumn{1}{c}{\#} &   & \multicolumn{1}{c}{[px]} & \multicolumn{1}{c}{[px]} & [mag] & [mag] & [mag] & [mag] \\
\hline
117	&	-	&	559.77	&	970.03	&	14.200	&	0.848	&	2.300	&	0.518	\\
118	&	F 9	&	1112.65	&	971.08	&	10.885	&	0.051	&	-1.015	&	-0.279	\\
119	&	F 23	&	867.23	&	971.34	&	13.824	&	1.039	&	1.924	&	0.709	\\
120	&	-	&	295.87	&	977.09	&	17.157	&	1.659	&	5.257	&	1.329	\\
121	&	-	&	427.70	&	980.10	&	16.458	&	1.318	&	4.558	&	0.988	\\
122	&	-	&	1054.97	&	984.20	&	16.442	&	1.416	&	4.542	&	1.086	\\
123	&	-	&	313.07	&	984.63	&	14.516	&	1.048	&	2.616	&	0.718	\\
124	&	EV 101	&	629.94	&	985.65	&	14.616	&	0.993	&	2.716	&	0.663	\\
125	&	-	&	947.79	&	986.76	&	15.805	&	1.239	&	3.905	&	0.909	\\
126	&	-	&	1110.64	&	994.16	&	15.072	&	1.298	&	3.172	&	0.968	\\
127	&	-	&	938.11	&	1000.83	&	16.425	&	1.493	&	4.525	&	1.163	\\
128	&	F 5	&	1313.96	&	1007.28	&	12.915	&	0.474	&	1.015	&	0.144	\\
129	&	-	&	49.92	&	1013.16	&	14.420	&	1.379	&	2.520	&	1.049	\\
130	&	-	&	1170.83	&	1015.02	&	17.255	&	1.962	&	5.355	&	1.632	\\
131	&	-	&	1377.98	&	1018.00	&	17.168	&	1.582	&	5.268	&	1.252	\\
132	&	-	&	347.13	&	1028.38	&	16.958	&	1.574	&	5.058	&	1.244	\\
133	&	EV 118	&	798.36	&	1041.11	&	15.004	&	0.921	&	3.104	&	0.591	\\
134	&	-	&	1777.01	&	1046.14	&	16.344	&	1.157	&	4.444	&	0.827	\\
136	&	T 52	&	168.39	&	1057.81	&	12.333	&	0.719	&	0.433	&	0.389	\\
137	&	-	&	764.65	&	1060.37	&	15.063	&	1.368	&	3.163	&	1.038	\\
138	&	T 53	&	373.26	&	1062.19	&	12.364	&	0.845	&	0.464	&	0.515	\\
139	&	F 14	&	1163.20	&	1064.25	&	9.908	&	0.011	&	-1.992	&	-0.319	\\
140	&	EV 117	&	600.77	&	1068.02	&	16.079	&	3.436	&	4.179	&	3.106	\\
141	&	EV 341	&	292.97	&	1072.18	&	14.870	&	0.673	&	2.970	&	0.343	\\
142	&	-	&	755.79	&	1072.22	&	16.298	&	1.206	&	4.398	&	0.876	\\
143	&	-	&	1142.33	&	1085.32	&	16.181	&	1.561	&	4.281	&	1.231	\\
144	&	F 25	&	957.10	&	1104.88	&	12.546	&	0.193	&	0.646	&	-0.137	\\
145	&	-	&	755.94	&	1107.58	&	14.528	&	0.920	&	2.628	&	0.590	\\
146	&	-	&	768.08	&	1109.97	&	14.881	&	1.161	&	2.981	&	0.831	\\
147	&	F 24	&	867.63	&	1117.11	&	11.401	&	0.099	&	-0.499	&	-0.231	\\
148	&	-	&	214.36	&	1119.42	&	15.477	&	1.107	&	3.577	&	0.777	\\
149	&	F 22	&	822.58	&	1123.93	&	12.344	&	0.586	&	0.444	&	0.256	\\
150	&	-	&	1137.82	&	1125.75	&	15.797	&	1.118	&	3.897	&	0.788	\\
151	&	F 18	&	1210.31	&	1127.16	&	13.414	&	0.895	&	1.514	&	0.565	\\
152	&	T 54	&	642.20	&	1143.42	&	12.343	&	0.574	&	0.443	&	0.244	\\
153	&	153	&	813.09	&	1160.87	&	15.900	&	1.043	&	4.000	&	0.713	\\
154	&	EV 131	&	460.58	&	1172.03	&	17.089	&	1.410	&	5.189	&	1.080	\\
155	&	-	&	1894.82	&	1181.04	&	15.942	&	1.326	&	4.042	&	0.996	\\
156	&	-	&	1723.72	&	1191.24	&	16.841	&	1.260	&	4.941	&	0.930	\\
157	&	-	&	1829.80	&	1196.28	&	16.631	&	1.169	&	4.731	&	0.839	\\
158	&	T 84	&	1921.94	&	1202.35	&	12.173	&	0.177	&	0.273	&	-0.153	\\
159	&	-	&	897.28	&	1208.26	&	16.722	&	1.459	&	4.822	&	1.129	\\
160	&	-	&	1733.69	&	1211.13	&	15.075	&	1.104	&	3.175	&	0.774	\\
161	&	-	&	1671.01	&	1211.58	&	14.758	&	0.802	&	2.858	&	0.472	\\
162	&	-	&	972.84	&	1214.98	&	17.186	&	1.603	&	5.286	&	1.273	\\
163	&	-	&	1067.84	&	1218.17	&	17.137	&	1.521	&	5.237	&	1.191	\\
164	&	T 5	&	499.27	&	1228.13	&	11.307	&	0.011	&	-0.593	&	-0.319	\\
165	&	-	&	766.29	&	1231.46	&	17.090	&	1.361	&	5.190	&	1.031	\\
166	&	EV 105	&	1428.35	&	1232.13	&	15.391	&	1.282	&	3.491	&	0.952	\\
167	&	F 3	&	1300.22	&	1238.68	&	10.329	&	0.281	&	-1.571	&	-0.049	\\
168	&	EV 119	&	746.33	&	1248.59	&	14.920	&	0.952	&	3.020	&	0.622	\\
169	&	-	&	454.11	&	1262.42	&	17.140	&	2.916	&	5.240	&	2.586	\\
170	&	F 4	&	1185.09	&	1262.77	&	12.900	&	0.702	&	1.000	&	0.372	\\
171	&	-	&	811.75	&	1266.25	&	17.083	&	1.403	&	5.183	&	1.073	\\
172	&	-	&	387.18	&	1267.98	&	16.583	&	1.102	&	4.683	&	0.772	\\
173	&	EV 120	&	650.21	&	1270.25	&	15.923	&	1.085	&	4.023	&	0.755	\\
174	&	-	&	354.41	&	1271.08	&	14.232	&	1.321	&	2.332	&	0.991	\\
175	&	EV 106	&	1578.70	&	1285.38	&	15.113	&	0.979	&	3.213	&	0.649	\\
\hline
\multicolumn{8}{l}{cross references according to: T ... Th\'e (1965), EV ... Evans (1978),} \\
\multicolumn{8}{l}{F ... Fitzgerald et al. (1978)} \\
\end{tabular}
\end{table*}

\begin{table*}
 \centering
\contcaption{Photometric data}
\begin{tabular}{@{}rcrrcccc@{}}
  \hline
no & ref & \multicolumn{1}{c}{X} & \multicolumn{1}{c}{Y} & V & $(B-V)$ & $M_V$ & $(B-V)_0$\\
\multicolumn{1}{c}{\#} &   & \multicolumn{1}{c}{[px]} & \multicolumn{1}{c}{[px]} & [mag] & [mag] & [mag] & [mag] \\
\hline
176	&	T 58	&	312.02	&	1289.90	&	12.425	&	0.339	&	0.525	&	0.009	\\
177	&	EV 121	&	578.36	&	1291.90	&	15.956	&	1.126	&	4.056	&	0.796	\\
178	&	-	&	213.07	&	1291.97	&	16.642	&	1.322	&	4.742	&	0.992	\\
179	&	T 83	&	1845.63	&	1296.59	&	9.581	&	0.020	&	-2.319	&	-0.310	\\
180	&	-	&	418.82	&	1306.86	&	16.830	&	1.286	&	4.930	&	0.956	\\
181	&	-	&	316.03	&	1310.79	&	17.028	&	1.935	&	5.128	&	1.605	\\
182	&	EV 104	&	1303.33	&	1319.82	&	15.217	&	1.000	&	3.317	&	0.670	\\
183	&	-	&	414.92	&	1321.43	&	14.659	&	0.837	&	2.759	&	0.507	\\
184	&	-	&	153.17	&	1340.29	&	15.929	&	1.199	&	4.029	&	0.869	\\
185	&	-	&	1935.35	&	1355.92	&	15.803	&	1.011	&	3.903	&	0.681	\\
186	&	-	&	1868.70	&	1357.27	&	15.862	&	1.005	&	3.962	&	0.675	\\
187	&	-	&	992.20	&	1359.94	&	16.896	&	1.131	&	4.996	&	0.801	\\
188	&	-	&	182.27	&	1360.00	&	16.524	&	1.882	&	4.624	&	1.552	\\
189	&	-	&	136.86	&	1361.14	&	14.973	&	0.814	&	3.073	&	0.484	\\
190	&	EV 103	&	1308.64	&	1367.22	&	16.275	&	1.087	&	4.375	&	0.757	\\
191	&	-	&	1678.35	&	1369.76	&	15.107	&	0.876	&	3.207	&	0.546	\\
192	&	-	&	986.65	&	1388.20	&	15.407	&	1.158	&	3.507	&	0.828	\\
193	&	-	&	1634.02	&	1391.19	&	15.577	&	0.950	&	3.677	&	0.620	\\
194	&	T 4	&	2011.94	&	1391.19	&	13.425	&	1.053	&	1.525	&	0.723	\\
195	&	T 82	&	1707.91	&	1404.01	&	13.222	&	0.963	&	1.322	&	0.633	\\
196	&	EV 102	&	1360.30	&	1409.28	&	14.729	&	0.802	&	2.829	&	0.472	\\
197	&	EV 514	&	1548.60	&	1412.10	&	14.088	&	0.633	&	2.188	&	0.303	\\
198	&	T 55	&	849.95	&	1418.74	&	12.827	&	0.627	&	0.927	&	0.297	\\
199	&	-	&	1237.07	&	1419.28	&	16.019	&	1.093	&	4.119	&	0.763	\\
200	&	-	&	1797.97	&	1435.41	&	15.211	&	0.929	&	3.311	&	0.599	\\
201	&	T 56	&	863.37	&	1439.33	&	13.828	&	0.893	&	1.928	&	0.563	\\
202	&	-	&	492.36	&	1443.38	&	16.411	&	1.590	&	4.511	&	1.260	\\
203	&	T 57	&	418.35	&	1454.55	&	10.690	&	0.190	&	-1.210	&	-0.140	\\
204	&	EV 127	&	904.69	&	1458.22	&	14.628	&	0.915	&	2.728	&	0.585	\\
205	&	-	&	42.27	&	1462.05	&	14.396	&	2.072	&	2.496	&	1.742	\\
206	&	-	&	663.53	&	1463.84	&	15.912	&	0.966	&	4.012	&	0.636	\\
207	&	-	&	1765.15	&	1465.00	&	16.533	&	1.117	&	4.633	&	0.787	\\
208	&	-	&	84.05	&	1474.30	&	16.125	&	1.286	&	4.225	&	0.956	\\
209	&	-	&	223.36	&	1476.16	&	16.487	&	1.421	&	4.587	&	1.091	\\
210	&	-	&	1455.78	&	1489.89	&	16.119	&	1.355	&	4.219	&	1.025	\\
211	&	EV 126	&	794.32	&	1490.92	&	15.320	&	0.881	&	3.420	&	0.551	\\
212	&	T 81	&	1630.73	&	1491.70	&	12.691	&	1.900	&	0.791	&	1.570	\\
213	&	EV 128	&	1552.35	&	1516.75	&	15.825	&	1.331	&	3.925	&	1.001	\\
214	&	-	&	1679.66	&	1526.65	&	15.995	&	0.996	&	4.095	&	0.666	\\
215	&	-	&	906.80	&	1527.87	&	16.612	&	1.157	&	4.712	&	0.827	\\
216	&	-	&	1479.14	&	1528.07	&	16.472	&	1.038	&	4.572	&	0.708	\\
217	&	-	&	374.19	&	1529.61	&	16.395	&	1.522	&	4.495	&	1.192	\\
218	&	-	&	495.27	&	1543.18	&	16.943	&	1.102	&	5.043	&	0.772	\\
219	&	-	&	156.18	&	1550.33	&	14.952	&	1.051	&	3.052	&	0.721	\\
220	&	-	&	1047.89	&	1558.93	&	16.671	&	2.027	&	4.771	&	1.697	\\
221	&	-	&	154.93	&	1558.95	&	16.539	&	1.190	&	4.639	&	0.860	\\
222	&	-	&	811.08	&	1559.72	&	16.820	&	1.531	&	4.920	&	1.201	\\
223	&	-	&	48.92	&	1570.62	&	15.104	&	1.995	&	3.204	&	1.665	\\
224	&	-	&	1102.75	&	1582.05	&	16.358	&	1.123	&	4.458	&	0.793	\\
225	&	-	&	1865.21	&	1586.25	&	15.578	&	0.886	&	3.678	&	0.556	\\
226	&	T 80	&	1494.10	&	1600.21	&	13.342	&	0.780	&	1.442	&	0.450	\\
227	&	-	&	1853.22	&	1609.97	&	16.519	&	0.584	&	4.619	&	0.254	\\
228	&	-	&	540.81	&	1629.21	&	15.321	&	0.923	&	3.421	&	0.593	\\
229	&	-	&	1003.84	&	1630.04	&	16.503	&	1.555	&	4.603	&	1.225	\\
230	&	-	&	485.73	&	1643.99	&	15.383	&	1.320	&	3.483	&	0.990	\\
231	&	-	&	1356.33	&	1681.92	&	16.414	&	1.181	&	4.514	&	0.851	\\
232	&	-	&	1167.04	&	1682.17	&	16.032	&	0.995	&	4.132	&	0.665	\\
233	&	-	&	1149.45	&	1684.80	&	15.258	&	0.929	&	3.358	&	0.599	\\
\hline
\multicolumn{8}{l}{cross references according to: T ... Th\'e (1965), EV ... Evans (1978),} \\
\multicolumn{8}{l}{F ... Fitzgerald et al. (1978)} \\
\end{tabular}
\end{table*}

\begin{table*}
 \centering
\contcaption{Photometric data}
\begin{tabular}{@{}rcrrcccc@{}}
  \hline
no & ref & \multicolumn{1}{c}{X} & \multicolumn{1}{c}{Y} & V & $(B-V)$ & $M_V$ & $(B-V)_0$\\
\multicolumn{1}{c}{\#} &   & \multicolumn{1}{c}{[px]} & \multicolumn{1}{c}{[px]} & [mag] & [mag] & [mag] & [mag] \\
\hline
234	&	-	&	1855.72	&	1696.06	&	16.588	&	1.236	&	4.688	&	0.906	\\
235	&	T 79	&	1558.33	&	1700.24	&	13.480	&	0.783	&	1.580	&	0.453	\\
236	&	-	&	1437.65	&	1701.57	&	14.253	&	0.779	&	2.353	&	0.449	\\
237	&	-	&	248.32	&	1702.52	&	17.885	&	1.871	&	5.985	&	1.541	\\
238	&	T 62	&	82.90	&	1703.12	&	12.942	&	0.828	&	1.042	&	0.498	\\
239	&	T 77	&	1876.77	&	1711.86	&	12.507	&	0.916	&	0.607	&	0.586	\\
240	&	-	&	1652.13	&	1714.77	&	16.039	&	1.002	&	4.139	&	0.672	\\
241	&	-	&	368.10	&	1720.81	&	15.411	&	1.304	&	3.511	&	0.974	\\
242	&	EV 125	&	787.44	&	1721.21	&	15.718	&	1.054	&	3.818	&	0.724	\\
243	&	EV 140	&	755.81	&	1732.73	&	14.914	&	1.529	&	3.014	&	1.199	\\
244	&	-	&	345.34	&	1734.22	&	14.181	&	2.003	&	2.281	&	1.673	\\
245	&	-	&	365.95	&	1750.00	&	15.843	&	1.045	&	3.943	&	0.715	\\
246	&	-	&	319.94	&	1752.38	&	16.910	&	2.713	&	5.010	&	2.383	\\
247	&	-	&	901.91	&	1752.96	&	16.854	&	1.268	&	4.954	&	0.938	\\
248	&	-	&	1639.58	&	1772.67	&	16.486	&	1.071	&	4.586	&	0.741	\\
249	&	EV 122	&	558.13	&	1775.02	&	14.052	&	0.754	&	2.152	&	0.424	\\
250	&	-	&	1255.14	&	1777.12	&	15.439	&	1.443	&	3.539	&	1.113	\\
251	&	-	&	1080.34	&	1784.27	&	15.613	&	0.895	&	3.713	&	0.565	\\
252	&	EV 123	&	665.71	&	1792.36	&	14.396	&	1.174	&	2.496	&	0.844	\\
253	&	EV 124	&	711.27	&	1793.36	&	15.343	&	1.170	&	3.443	&	0.840	\\
254	&	-	&	72.90	&	1797.44	&	15.741	&	1.221	&	3.841	&	0.891	\\
255	&	-	&	409.24	&	1799.02	&	16.521	&	1.188	&	4.621	&	0.858	\\
256	&	-	&	1186.89	&	1799.37	&	15.286	&	0.919	&	3.386	&	0.589	\\
257	&	T 71	&	1253.82	&	1808.40	&	12.794	&	0.867	&	0.894	&	0.537	\\
258	&	-	&	2016.97	&	1814.63	&	14.563	&	1.018	&	2.663	&	0.688	\\
259	&	-	&	125.07	&	1833.07	&	16.374	&	1.533	&	4.474	&	1.203	\\
260	&	-	&	846.68	&	1838.83	&	16.057	&	1.238	&	4.157	&	0.908	\\
261	&	T 78	&	1713.86	&	1844.37	&	12.375	&	0.779	&	0.475	&	0.449	\\
262	&	T 63	&	193.20	&	1848.21	&	12.487	&	0.671	&	0.587	&	0.341	\\
263	&	-	&	915.88	&	1852.10	&	17.024	&	2.334	&	5.124	&	2.004	\\
264	&	-	&	106.20	&	1859.70	&	15.449	&	0.997	&	3.549	&	0.667	\\
265	&	-	&	1820.86	&	1863.28	&	15.305	&	1.167	&	3.405	&	0.837	\\
266	&	T 64	&	290.91	&	1864.59	&	11.234	&	0.769	&	-0.666	&	0.439	\\
267	&	T 10	&	574.28	&	1865.66	&	10.048	&	-0.016	&	-1.852	&	-0.346	\\
268	&	-	&	1769.05	&	1868.72	&	15.043	&	1.169	&	3.143	&	0.839	\\
269	&	-	&	1606.88	&	1870.61	&	16.628	&	1.209	&	4.728	&	0.879	\\
270	&	EV 133	&	1302.32	&	1875.69	&	16.390	&	2.501	&	4.490	&	2.171	\\
271	&	-	&	783.40	&	1880.84	&	13.844	&	1.246	&	1.944	&	0.916	\\
272	&	-	&	1376.96	&	1894.95	&	15.457	&	2.191	&	3.557	&	1.861	\\
273	&	T 70	&	985.36	&	1912.30	&	13.074	&	0.502	&	1.174	&	0.172	\\
274	&	-	&	2016.04	&	1921.81	&	14.643	&	0.912	&	2.743	&	0.582	\\
275	&	T 65	&	443.72	&	1927.54	&	10.629	&	1.296	&	-1.271	&	0.966	\\
276	&	-	&	1021.39	&	1928.14	&	15.421	&	0.823	&	3.521	&	0.493	\\
277	&	-	&	807.00	&	1932.98	&	15.108	&	1.265	&	3.208	&	0.935	\\
278	&	-	&	995.31	&	1948.07	&	16.408	&	2.347	&	4.508	&	2.017	\\
279	&	-	&	1194.13	&	1957.44	&	15.576	&	1.264	&	3.676	&	0.934	\\
280	&	-	&	1201.84	&	1975.30	&	14.674	&	0.858	&	2.774	&	0.528	\\
281	&	T 69	&	876.93	&	1980.01	&	12.777	&	0.896	&	0.877	&	0.566	\\
282	&	-	&	1948.21	&	1983.77	&	15.008	&	0.865	&	3.108	&	0.535	\\
283	&	-	&	176.35	&	1987.67	&	15.584	&	1.110	&	3.684	&	0.780	\\
284	&	-	&	716.69	&	1991.23	&	14.702	&	2.120	&	2.802	&	1.790	\\
285	&	EV 316	&	1754.04	&	1992.89	&	14.539	&	0.739	&	2.639	&	0.409	\\
286	&	-	&	35.14	&	2016.29	&	16.176	&	1.735	&	4.276	&	1.405	\\
287	&	-	&	1224.00	&	2018.87	&	15.302	&	1.108	&	3.402	&	0.778	\\
288	&	-	&	1162.16	&	2022.13	&	14.807	&	0.997	&	2.907	&	0.667	\\
289	&	-	&	424.21	&	2022.17	&	16.411	&	1.878	&	4.511	&	1.548	\\
\hline
\multicolumn{8}{l}{cross references according to: T ... Th\'e (1965), EV ... Evans (1978),} \\
\multicolumn{8}{l}{F ... Fitzgerald et al. (1978)} \\
\end{tabular}
\end{table*}

\label{lastpage}

\end{document}